# Element-specific ultrafast lattice dynamics in monolayer WSe$_2$

H. Jung[1,2,*], S. Dong[2,a], D. Zahn[2,b], T. Vasileiadis[2,c], H. Seiler[2,d], R. Schneider[3], S. Michaelis de Vasconcellos[3], V. C. A. Taylor[2], R. Bratschitsch[3], R. Ernstorfer[1,2], Y. W. Windsor[1,2,†]

[1] *Institute for Optics and Atomic Physics, Technical University Berlin, Strasse des 17. Juni 135, 10623 Berlin, Germany*
[2] *Department of Physical Chemistry, Fritz Haber Institute of the Max Planck Society, Faradayweg 4-6, 14195 Berlin, Germany*
[3] *Institute of Physics and Center for Nanotechnology, University of Münster, Heisenbergstr. 11, 48149 Münster, Germany*

[a] *Present address: Institute of Physics, Chinese Academy of Sciences, Beijing, 100190, China*

[b] *Present address: Fraunhofer Institute for Electronic Microsystems and Solid State Technologies (EMFT), Hansastraße 27d, 80686 München, Germany*

[c] *Present address: Faculty of Physics, Adam Mickiewicz University, Poznań, ul. Uniwersytetu Poznańskiego 2, 61-614 Poznań, Poland*

[d] *Present address: Institute of Physics, Freie Universität Berlin, Arnimallee 14, 14195 Berlin, Germany*

[*] hyein.jung@tu-berlin.de

[†] windsor@tu-berlin.de

ABSTRACT:

We study monolayer WSe$_2$ using ultrafast electron diffraction. We introduce an approach to quantitatively extract atomic-site-specific information, providing an element-specific view of incoherent atomic vibrations following femtosecond excitation. Via differences between W and Se vibrations, we identify stages in the nonthermal evolution of the lattice. Combined with a calculated phonon dispersion, this element specificity enables us to identify a long-lasting overpopulation of specific optical phonons, and to interpret the stages as energy transfer processes between specific phonon groups. These results demonstrate the appeal of resolving element-specific vibrational information in the ultrafast time domain.

Van der Waals bonded materials consist of weakly bonded atomic layers, facilitating the practical synthesis of samples at thicknesses down to a single atomically-thin crystalline layer. Such dimensional constraints can strongly alter bulk properties, such as an indirect band gap becoming direct [1–3]. The availability of such monolayers fuels the realization of ultrathin single-crystalline monolayer-monolayer heterostructures, where the combination of functionally distinct monolayers can enable new functional properties [4,5]. Transition metal dichalcogenides (TMDCs) are such materials, which are predicted to play a central role in next-generation optoelectronics due to their infrared-to-visible bandgaps. [6,7]

For future applications, understanding the non-equilibrium behavior of monolayer TMDCs is paramount. To date, femtosecond studies have revealed a variety of nonequilibrium electronic phenomena [8–13]. The ultrafast excitations in these studies are also expected to affect the lattice via electron-phonon coupling (EPC). EPC is expected to initiate a cascade of phonon-phonon scattering processes [14,15], through which the phonon population evolves to a new thermalized equilibrium state. Understanding sub-picosecond phenomena within the lattice is of particular interest in monolayers due to their constrained dimensions, which have been recently shown to affect EPC [16].

Beyond monolayer TMDCs, studying the sub-picosecond evolution of phonons elucidates the coupling between vibrational modes, holding great potential for nanoscale devices, e.g. for engineering heat flow and novel control pathways [17,18]. Despite this, experimental reports about the ultrafast evolution of phonon populations, in particular through nonthermal states (those that disobey Bose-Einstein statistics, meaning that temperatures cannot describe them), are rare. Femtosecond electron- and X-ray scattering techniques can access such information and have been reported on both metals [19–25] and semiconductors. [26–30] Ultrafast Bragg diffraction (elastic scattering), is particularly appealing because it provides quantitative real-space information about changes in the atomic arrangement and the atomic vibrations, including incoherent vibrations through the Debye-Waller effect. [26,31] Fully resolving such motions is particularly advantageous when studying multi-element compounds, as it can elucidate different ions' roles in the ultrafast vibrational response to photoexcitation. This has proven to be challenging, and studies often resort to "effective" material-averaged information [28,32–34], or to comparison with theoretical simulations of such motion. [27,34,35]. In fact, to our knowledge, there are presently no reports in literature in which crystal structures were fully resolved in the ultrafast time domain (apart from single elements).

A recent ultrafast electron diffraction (UED) study on NiO demonstrated progress toward element specificity [36]. This was achieved by employing tabulated equilibrium Debye-Waller factors of each element and identifying photo-induced variations from them. Here we advance beyond this approach, by independently determining element-specific Debye-Waller factors as functions of time, purely from the experimental data only. This provides a real space picture of atomic motion and can highlight the role of individual atoms, an approach that is complementary to the momentum resolution obtained by inelastic scattering [16,33,37–39]. For example, V-dominated incoherent vibrations around ~3 THz were theoretically suggested as a key ingredient in the ultrafast structural phase transition in $VO_2$ [35], which has been the topic of several ultrafast diffraction studies [40–42].

We present a UED study of monolayer WSe$_2$. We record hundreds of Bragg reflections, enabling us to disentangle and quantitatively determine the incoherent vibrations of all W and of Se atoms on femtosecond time scales. This allows us to identify independent trends for each element, elucidating different stages within the phonon-phonon thermalization process. By combining these with an element-specific breakdown of the phonon dispersion, we relate the experimental results to a phonon picture and identify a prolonged overpopulation of low-energy optical phonons.

A monolayer WSe$_2$ sample was prepared by micromechanical exfoliation of a bulk crystal and subsequently transferred to a 10 nm thick amorphous Si$_3$N$_4$ membrane substrate by an all-dry transfer technique [43]. The substrate is a 100 × 100 μm$^2$ window, which is approximately the size of the probe beam, (see Fig. 1a) within a silicon frame, such that the entire sample area is probed. Fig. 1b presents a photoluminescence spectrum taken from the sample, exhibiting a pronounced A-exciton resonance at 1.66 eV, as expected from monolayer WSe$_2$ [44]. Fig. 1c presents a sketch of the 4 kHz room-temperature UED experiment. [45] A photon pulse with $hv = 1.65$ eV (751 nm) optically excites the sample, followed by a 75 KeV electron pulse arriving at a variable time delay, *t*. The electrons arrive normal to the sample surface and are transmitted through it, producing a diffraction pattern on a 2D detector, as in Fig. 1d (a faint Scherrer ring is caused by the substrate). The excited carrier density generated by the incident 1.4 mJ cm$^{-2}$ photon pulse is (4.3 ± 0.1) × 10$^{14}$ carriers/cm$^{-2}$ (see supplement), which is much higher than the reported Mott density (~10$^{13}$ cm$^{-2}$). [46,47] This suggests that the majority of excited carriers behave as quasi-free carriers, so excitonic effects are not considered herein, despite the agreement between the photon energy and the exciton resonance.

The observed Bragg reflections provide in-plane sensitivity and are denoted with Miller indices (*hk0*). Each Bragg reflection is described by a scattering vector ***q*** (as indicated in Fig. 1d). We define the scattering vector as $|\boldsymbol{q}| = 4\pi\lambda^{-1}\sin\theta$ (λ is the electrons' de Broglie wavelength and θ is the Bragg angle). For hexagonal monolayer WSe$_2$, ***q*** satisfies $|\boldsymbol{q}|^2 = (2\pi/a)^2 \left[\frac{4}{3}(h^2 + k^2 + hk)\right]$, with $a = 3.297$ Å [48]. Note that Bragg reflections with different Miller indices can have the same scattering vector length $|\boldsymbol{q}|$.

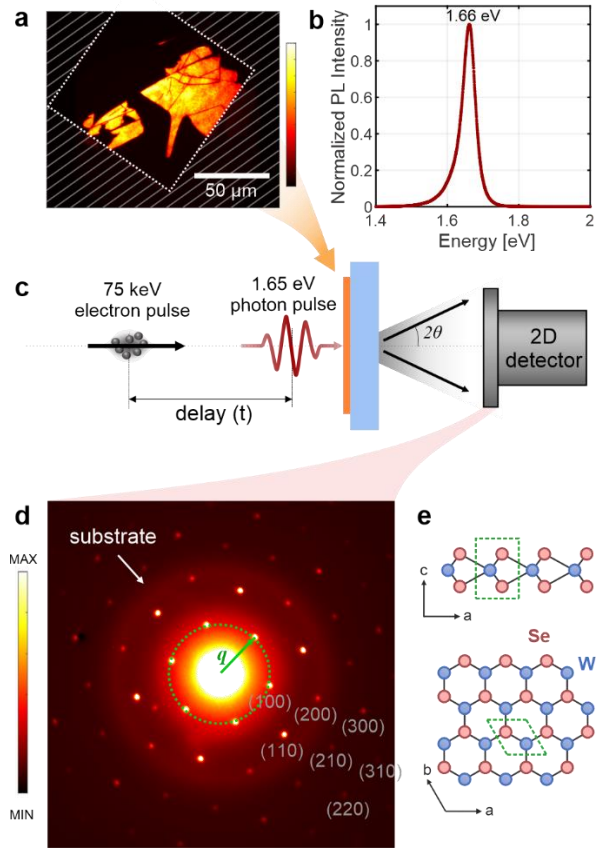

**Fig. 1** (a) A photoluminescence image of the sample on the 100×100 μm² membrane window. The hatched area corresponds to thick silicon. (b) Photoluminescence spectrum of the sample. (c) Schematic experimental setup of ultrafast electron diffraction: an optical pulse excites the sample at nearly normal incidence. After a variable time delay an electron pulse transmits through the sample at normal incidence. The scattered electrons produce a diffraction pattern on a 2D detector at each time delay. (d) An example of such a diffraction pattern, exhibiting Bragg spots from the monolayer WSe₂ and Scherrer rings from the substrate. One group of reflections with the same $|q|$ is indicated. (e) Side- and top-down views of the WSe₂ monolayer structure, with the unit cell highlighted.

We collect diffraction patterns for several pump-probe delays. From these we extract the intensities of the Bragg reflections, which are described analytically by: [49,50]

$$I(\boldsymbol{q},t) \propto \left| \sum_j f_j^{(e)}(q) \tau_j(t,\boldsymbol{q}) \exp(i\boldsymbol{q} \cdot \boldsymbol{r}_j) \right|^2. \quad (1)$$

Here we sum over all three atoms in the monolayer unit cell (Fig. 1e): $f_j^{(e)}(q)$ are tabulated electron scattering factors of Se or W atoms [51,52], $r_j$ is the relative equilibrium position of the $j$-th atom, and $\tau_j$ is its temperature factor (explained below). Since we do not observe a coherent response within our time resolution, we do not consider changes to $r_j$, and subsequently Eq. (1) yields the same intensity for all (*hk0*) combinations that produce the same $|q|$. We therefore average the intensities of all reflections that share the same $|q|$. Fig. 2a presents these averaged intensities, normalized by their unperturbed values.

Photoexcitation suppresses all intensities due to the growth of the incoherent phonon population. For a microscopic understanding, we use Eq. (1) to interpret the results in Fig. 2a, such that the temperature factors $\tau_j$ encode all dynamics. These express the Debye-Waller effect, through which Bragg reflections are attenuated with increased atomic vibrations (e.g. when heated). These vibrations are quantified using a time-averaged matrix $\mathbf{U}^{(j)}$, which describes the mean-squared displacement (MSD) of atom $j$ around its equilibrium position, along different spatial directions. The MSD encodes all incoherent atomic vibrations, which are a superposition of all populated phonon modes. [49,50]. The temperature factor $\tau_j$ is expressed as: [50]

$$\tau_j = \exp\left(-\tfrac{1}{2}\left(U_{11}^{(j)}h^2 a^{*2} + U_{22}^{(j)}k^2 b^{*2} + 2U_{12}^{(j)} h a^* k b^*\right)\right), \qquad (2)$$

where $a^*$ and $b^*$ are the reciprocal lattice vector lengths. $U_{11}^{(j)}$, $U_{22}^{(j)}$, and $U_{12}^{(j)}$ are elements of $\mathbf{U}^{(j)}$ along different in-plane directions (this experiment is not directly sensitive to out-of-plane MSD contributions). The in-plane local environment of the atoms in monolayer WSe$_2$ is the same as that of bulk. We therefore consider the same in-plane symmetry constraints on $\mathbf{U}^{(j)}$ for space group #194 (with $l = 0$). For both the W and Se crystallographic sites, these dictate that $U_{11}^{(j)} = U_{22}^{(j)} = 2U_{12}^{(j)} \equiv U^{(j)}$ [49]. The in-plane MSD of each element is therefore described by a single parameter, $U^{(j)}$, and Eq. (2) reduces to

$$\tau_j = \exp\left(-\tfrac{1}{2} U^{(j)} a^{*2}(h^2 + k^2 + hk)\right). \qquad (3)$$

Note that use of $\tau_j$ is valid for describing nonthermal states, and its name "temperature factor" is purely common nomenclature. The effect of photoexcitation is now expressed in the delay-dependent change of each element's MSD, $\Delta U^{(j)}(t)$. We therefore consider $U^{(j)}$ as the sum $U^{(j)}(t) = U_0^{(j)} + \Delta U^{(j)}(t)$, in which $U_0^{(j)}$ is the unperturbed MSD. Finally, by inserting $U^{(j)}$ into $\tau_j$ (Eq. (3)), and then inserting $\tau_j$ into Eq. (1), we reach a set of microscopic equations for each unique $I(\mathbf{q}, t)$ that depend only on two dynamic variables: $\Delta U^{(W)}(t)$ and $\Delta U^{(Se)}(t)$, which are independent of $\mathbf{q}$. This means that each curve in Fig. 2a can be described by the evolution of these two quantities. Note that since the two Se atoms in the monolayer unit cell share the same equilibrium Wyckoff position, we assume that $U^{(Se)}(t)$ is the same for both. This assumption neglects differences between them due to the substrate's proximity to one side only.

To implement this, we employed a two-level routine that fits all observed intensities in Fig. 2a to this equation. First, for all delays, all $I(\mathbf{q}, t)$ must share the same unperturbed MSD values from each element, $U_0^{(W)}$ and $U_0^{(Se)}$. Second, all $I(\mathbf{q}, t)$ at a *given* delay share the same pair of $\Delta U^{(W)}$ and $\Delta U^{(Se)}$ values. The optimal values obtained for $U_0^{(Se)}$ and $U_0^{(W)}$ were $6.3 \times 10^{-3}$ Å$^2$ and $3.4 \times 10^{-3}$ Å$^2$, respectively, in agreement with experimental values reported for a closely related material, TiSe$_2$ [53] as well as with calculated values for monolayer WSe$_2$. [27] (See supplement).

Fig. 2b presents the resulting evolution of $\Delta U^{(W)}(t)$ and $\Delta U^{(Se)}(t)$. This represents an element-specific perspective of the evolution of incoherent atomic vibrations. By inspecting the changes in trends exhibited by $\Delta U^{(W)}$ and $\Delta U^{(Se)}$ (Fig. 2b), we observe three distinct stages:

I. $t < 2$ ps : both W and Se exhibit a rapid increase in MSD.
II. $2 < t < 4$ ps : opposite trends emerge, as the MSD of Se keeps growing, while that of W decreases rapidly.
III. $t > 4$ ps : all MSD evolution processes slow down, as the MSD of Se decreases, while the rapid decrease of W's MSD slows down and is eventually overtaken by slight growth at ~20 ps. (see supplement for dynamics on a linear time delay scale)

Variations in MSD can be described as an evolution of the phonon population (see supplement). According to Ref. [14], when semiconductors are photoexcited and undergo electron cooling, intravalley carrier scattering with smaller momentum and higher energy tends to occur, favoring stronger coupling to optical phonons near the Γ point, than to other phonon groups. This is supported by calculations on bulk $WSe_2$ [26] and is also consistent with reports on other monolayer crystals, e.g. graphene and $MoS_2$ [15,54]. It is then likely that phonon thermalization in monolayer $WSe_2$ starts with overpopulating higher energy optical modes near the Γ point, followed by populating lower energy modes instead via phonon-phonon scattering. The continuous increase in $\Delta U^{(Se)}$ compared to the decrease of $\Delta U^{(W)}$ in stage II suggests a preferential growth in the population of phonon modes dominated by Se vibrations, and cannot agree with an increase in lattice temperature. As such, stage II is direct evidence that the lattice is in a nonthermal state.

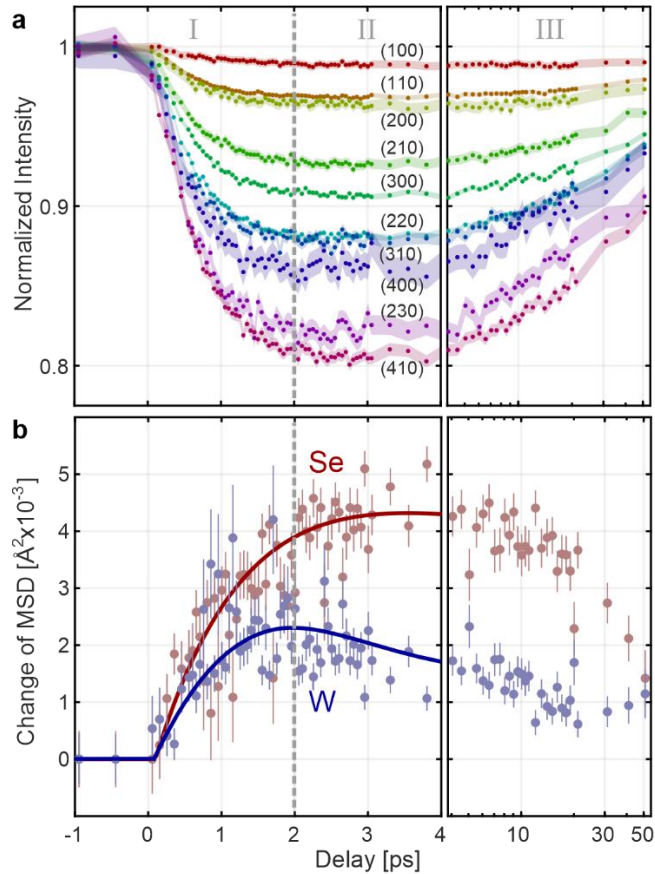

**Fig. 2** (a) Normalized Bragg reflection intensities as functions of pump-probe delay. Each curve represents the average of all reflections with the same scattering vector length and is denoted by a representative (*hk*0) combination. (b) The change in atomic mean-squared displacement (MSD) of each element ($\Delta U^{(W)}$ and $\Delta U^{(Se)}$), as extracted from all data in panel (a) using Eq. (1). The error bars are the fit standard deviations. The solid lines are drawn lines serving as guides to the eye. Time-ranges labelled I, II, and III indicate stages in the evolution of MSDs based on changes in their trends.

For further interpretation in terms of phonons, we conducted a density functional perturbation theory (DFPT) calculation using the QUANTUM ESPRESSO package. [55] A fully relaxed atomic structure was adapted using the plane-wave self-consistent field program. The electronic ground state is evaluated using a 10×10×1 mesh (convergence threshold: 1×10$^{-9}$ Ry/Bohr). Subsequently, the phononic structure is calculated for the dynamical matrices on a 8×8×1 q-grid (accurate consistency threshold: $1 \times 10^{-14}$ Ry/Bohr).

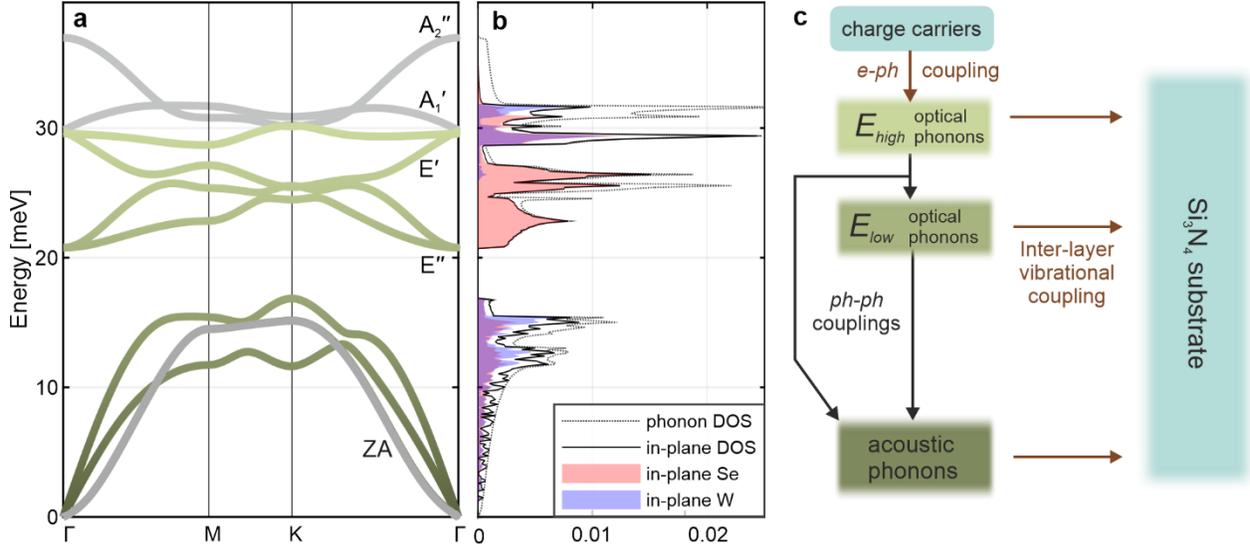

**Fig. 3** (a) Calculated phonon dispersion of monolayer WSe$_2$. Gray-colored bands are polarized out of plane, so the experiment is nearly insensitive to them (Eq. (2)). (b) Phonon density of states (DOS): total (dotted line), total in-plane (solid line), and breakdown of in-plane DOS by atom type (shades of red and blue). (c) Energy flow diagram illustrating phonon thermalization. The color scheme relates to the bands in (a).

Fig. 3a presents the calculated phonon dispersion of monolayer WSe$_2$. Out-of-plane polarized phonon modes are depicted in gray. Our experiment is nearly insensitive to them because they have very small in-plane contributions. Fig. 3b presents the corresponding phonon density of states (DOS; dotted line). The DOS was subdivided twice: polarization dependence (*in-* and *out-of-plane*) and elemental dependence (W or Se vibrations). Of these, Fig. 3b presents *in-plane* components of the DOS: total in-plane (solid line), and element-specific (shadings). We find that in-plane polarized states account for most of the DOS (except at the highest energies). These are vibrations to which our experiment is directly sensitive (Eq. (2)).

The key result of this calculation is the element-specific breakdown of in-plane vibrations, as it relates to element-resolved experimental results in Fig. 2b. By combining the two, we can interpret which phonons are dominant in each stage of thermalization. The primary observation from Fig. 3b is that the in-plane vibrations of Se dominate the lower-energy range of optical phonons ($E_{low} \approx 21$ to 27 meV), while W vibrations are notably absent. In other energy ranges, including the acoustic branches, W vibrations contribute similarly to Se.

Based on Fig. 3b, we now further interpret the stages in the evolution of the MSD curves (Fig. 2b). To guide this, we present an energy flow diagram (Fig. 3c). In stage I, the similar increase in $\Delta U^{(Se)}$ and $\Delta U^{(W)}$ suggests an initial growth in occupation of the high-energy optical branches ($E_{high} > \sim 27$ meV), particularly in states near the Γ point (low momentum), where Se and W vibrations are similar (Fig. 3b). This occurs due to energy flow from the excited carriers via EPC. In stage II, the dominance of Se vibrations over W (Fig. 2b) suggests a growth in the $E_{low}$ optical phonons (dominated by Se, Fig. 3b) through a loss of $E_{high}$ phonons via phonon-phonon scattering. The data suggest that a preferentially high population of these Se-dominated phonons persists for several ps. Lastly, to interpret the MSD behavior in stage III, we note that an unambiguous observation here is that the intensity of all Bragg reflections recovers (Fig. 2a), indicating a weakening of the Debye-Waller effect. This is typically interpreted as a result of energy flow away from the excited sample volume as part of a slow thermalization of the whole system back to the original equilibrium. Indeed, in stage III the W vibrations approach those of Se, which drop faster (Fig 2b; see also supplement, Fig. S2). This can indicate a growth in the acoustic phonon population because both elements contribute similarly to them (Fig. 3b), unlike the $E_{low}$ phonons.

Critically, our data suggests another energy flow process. This is because the highest MSD values are observed at the transition from stage II to stage III. If energy is indeed flowing from the $E_{low}$ phonons into acoustic phonons, we expect a further increase in MSD, because MSD typically scales as $E^{-1}$ [49,50], implying that acoustic phonons contribute most to MSD amplitudes. Furthermore, energy conservation would dictate that a high-energy phonon creates multiple low-energy phonons, each contributing more to MSD. However, as the Se-dominated signature decreases, so does the overall MSD, which is inconsistent with this expectation. We consider two possibilities.

The first is energy flow through vibrational coupling to the substrate, which is not photoexcited because of its large electronic band gap. Furthermore, the phonon dispersion in $Si_3N_4$ greatly exceeds the energy range in Fig. 3a, and within this range, its bands are dense with no energy gaps. [56] This suggests that coupling can occur from any $WSe_2$ optical phonons, diverting energy away from the sample before it reaches the acoustic phonons, effectively "shunting" their expected population growth (Fig. 3c). The recovery of all Bragg reflections (Fig. 2a stage III) supports this explanation, as does an observed attenuation of the substrate's Scherrer rings (stage II; see supplement), indicating that such vibrational coupling does indeed occur. This implies that "acoustic phonon shunting" may serve as a novel approach towards heat management in future nanoscale devices, with one layer conducting charge and another conducting heat.

Another possibility is preferential generation of out-of-plane polarized phonons, to which our experiment is largely insensitive. These particularly include the ZA modes, which at small momenta near the Γ point disperse at lower energies than all other modes. [15,57] The recovery at the later delays in Fig. 2a would then reflect a weakening of in-plane vibrations, but not of the total vibrations, as the system continues to thermalize. We note that ultrafast occupation of out-of-plane modes in monolayers is debated. [15,27,54]

In summary, we used ultrafast electron diffraction to probe photoinduced lattice dynamics in monolayer $WSe_2$. Each element's amplitude of incoherent vibrations was quantitatively extracted as a function of delay, producing an element-specific view of the vibrational response to photoexcitation. From differences between the atom species' vibrational responses, we identify stages in phonon-phonon thermalization. By combining this with an elemental breakdown of the phonon dispersion, we present a scenario in which an initial high-energy phonon population is generated, followed by the preferential population of low-energy optical phonons, which persists for several ps. This is followed by a rapid recovery of all Bragg intensities, which disagrees with the arrival at an elevated thermal phonon distribution, because an increase of acoustic phonons is not observed. We discuss two explanations: energy flow from the optical phonons directly to the substrate, or to ZA phonons, to which we are insensitive. The former explanation may serve as a route for heat management in nanoscale devices by shunting acoustic phonon generation. Beyond the specific case of WSe$_2$, the approach to time-resolved diffraction demonstrated here provides an element-resolved view of ultrafast phonon-phonon interactions. This can be readily applied to other materials, particularly when certain atoms are associated with specific processes or properties like magnetism. With further computation, phonon-branch-specificity is also conceivable. This real-space view is complementary to the momentum-resolved view obtained through diffuse scattering. Both serve to elucidate vibrational energy flow, which can ultimately lead to more efficient thermal management and improved performance in electronic devices.

This work received funding from the DFG within Transregio TRR 227 Ultrafast Spin Dynamics (Project A10) and the Priority Program SPP 2244 (project 443366970). Funding was also received from the Max Planck Society and the European Research Council (ERC) under the European Union's Horizon 2020 research and innovation program (Grant Agreement No. ERC-2015-CoG-682843). V.C.A.T. acknowledges financial support from the Alexander von Humboldt Foundation. H. S. acknowledges support from the Swiss National Science Foundation under Grant No. P2SKP2.184100.